\documentstyle[12pt]{article}

\begin{document}
\begin{flushright}
July, 1997
\end{flushright}
\vspace{2mm}
\begin{center}
\large{Chiral Symmetry and the Nucleon Structure Functions}
\end{center}
\vspace{6mm}
\begin{center}
M.~Wakamatsu and T.~Kubota
\end{center}
\vspace{2mm}
\begin{center}
Department of Physics, Faculty of Science, 
\end{center}
\begin{center}
Osaka University, 
\end{center}
\begin{center}
Toyonaka, Osaka 560, JAPAN
\end{center}
\vspace{7mm}
\begin{flushleft}
\end{flushleft}
\vspace{5mm}
\paragraph{Abstract :}

\ \ The flavor asymmetry of the sea quark distribution as well as
the unexpectedly small quark spin fraction of the nucleon are
two outstanding discoveries recently made in the physics of
deep-inelastic structure functions. We evaluate here the
corresponding quark distribution functions within the framework of
the chiral quark soliton model, which is an effective quark model
of baryons maximally incorporating the most important feature of
low energy QCD, i.e. the chiral symmetry and its spontaneous
breakdown. It is shown that the model can explain qualitative
features of the above-mentioned nucleon structure functions within
a single framework, thereby disclosing the importance of chiral
symmetry in the physics of high energy deep-inelastic scatterings.

\newpage
\section{Introduction}

\ \ \ Though it is certainly true that high energy deep-inelastic
scatterings provide us with the best testing ground for
perturbative QCD, the physics behind it is highly nonperturbative.
In fact, the perturbative QCD can describe only the evolution of
structure functions from one energy scale to the other, so that
for determining structure functions themselves one must retreat
to a semi-theoretical fitting procedure of empirical cross
sections at a certain scale [1].
In recent years, two epoch-making
findings have been made concerning the nucleon structure functions.
The first concerns the EMC measurement of the spin structure of the
proton through deep-inelastic scatterings of polarized muons on
polarized proton, which sometimes has been advertised as the
so-called ``nucleon spin crisis'' [2].
Another interesting observation
made by the NMC group is the flavor asymmetry of the sea quark
distributions in the nucleon [3]. As for this second observation,
a widely-accepted explanation is based on the pion cloud effects,
which may in turn be interpreted as a manifestation of chiral
symmetry of QCD in high energy observables [4-7].
The first observation, i.e. the nucleon spin problem is a little
more mysterious than the second. The chiral symmetry may play an
important role also here [8],
or it may be explained by some other nonperturbative effects like
gluon polarization dictated by the $U_A (1)$ anomaly
of QCD [9,10]. 

At any rate, these two remarkable findings have reminded us of a
quite plain fact : we absolutely need some theoretical device with
which we can evaluate structure functions in a nonperturbative way.
Unfortunately, nonperturbative treatment of QCD is quite involved.
Then, lattice gauge theory is widely believed to be the most
promising tool for it [11]. In the long run, it may be true.
Confining to the present status, however, it suffers from quite
a few obstacles mainly arising from the limitation of the computer
ability. Among others, what seems most serious to us is the use
of the so-called ``quenched approximation'', the validity of which
is strongly suspected when applied to the physics of light
quark sector that we are just interested in. At least, it
seems very unlikely that a reliable estimation of sea quark
distributions is feasible under this approximation.
Here is a place where effective theories of QCD can play unique
and important roles. Naturally, there exist quite a few effective
models of baryons, and some of them were already applied for
evaluating nucleon structure functions [12-14].
Here, we emphasize one
absolute superiority of the chiral quark soliton model (CQSM)
over the other, i.e. its ability to solve the nucleon bound state
problem nonperturbatively with full inclusion of the deformed
Dirac sea quarks in addition to three valence quarks [15-17].
Undoubtedly, this must be a unique advantage not shared by other
models of baryons when discussing the physics of the
Gottfried sum [6].

Several group have already attempted to calculate nucleon structure
functions within the CQSM or the NJL soliton model. For instance,
Weigel et al. evaluated the isovector unpolarized distribution
functions as well as some other distribution functions [18].
However, their calculation is incomplete in respect that the
effect of afore-mentioned Dirac sea quarks are totally neglected.
On the other hand, both of Diakonov et al. [19,20]
and Tanikawa and Saito [21] carried out more
consistent calculation including the deformed Dirac sea quarks,
but by confining themselves to the isosinglet unpolarized and
isovector polarized distribution functions, which have values at
the leading order of $1 / N_c$ expansion (or the expansion in the
collective rotational velocity $\Omega$ of the hedgehog soliton).
Unfortunately, the isovector unpolarized and isoscalar polarized
distribution functions both vanish at this order. Then, for
discussing interesting physics like the Gottfried sum or the
quark spin content of the nucleon, one must go ahead to the
next-to-leading order contribution in $\Omega$, which turns out
to be quite an involved task. The present paper is the first
report of such an investigation.

The plan of the paper is as follows. In sect.2, we shall derive
general formulas for evaluating quark distribution functions of
the nucleon on the basis of the path-integral formulation of the
chiral quark soliton model (CQSM). The results of the numerical
calculation are given and analyzed in sect.3. We then summarize
our results in sect.4.

\vspace{4mm}
\section{Theory of quark distribution functions}

\ \ \ We start with the light-cone expression for the
quark distribution function [12-14] :
\begin{equation}
   q (x) \ \ = \ \ \frac{1}{4 \pi} \,\,\int_{-\infty}^\infty
   \,\,d z^{-} \,\,e^{- \,i \,x \,P^+ \,z^-} \,\,
   {\langle N (P) \,| \, \bar{\psi} (z) \,\Gamma \,\psi(0) \,| \,
   N (P) \rangle} \,\, {|}_{z^+ = 0,\,z_\perp = 0} \,\, ,
\end{equation}
with the definition of the standard light-cone coordinates
$z^{\pm} = (z^0 \pm z^3) / \sqrt{2}$. The nucleon state is
normalized here as
\begin{equation}
   \langle N (P') \,| \,N(P) \rangle \ \ = \ \ 2 \,P^0 \,\,\delta^3
   (\mbox{\boldmath $P$}^\prime - \mbox{\boldmath $P$}) \,\, ,
\end{equation}
with $P$ the nucleon 4-momentum, while
$x = - \,q^2 / (2 P \cdot q)$ is the Bjorken variable with $q$
the 4-momentum transfer to the nucleon. We are to take
$\Gamma = \gamma^{+} \tau_3$ with $\gamma^+ = (\gamma^0 + 
\gamma^3) / \sqrt{2}$ for the isovector unpolarized
distribution function $u(x) - d(x)$, while $\Gamma = 
\gamma^{+} \gamma_5$ for the isosinglet polarized one
$\Delta u(x) + \Delta d(x)$.
We recall the fact that, extending the definition of
distribution function $q(x)$ to the interval
$-1 \le x \le 1$, the relevant antiquark distributions are
given as [19]
\begin{eqnarray}
   \bar{u} (x) - \bar{d} (x) \ \ &=& - \,\,[ \,u(-x) - d(-x) \,]
   \,\, ,\\
   \Delta \bar{u}(x) + \Delta \bar{d} (x)  &=& \ \ 
   \Delta u(-x) + \Delta d(-x)
   \,\, ,
\end{eqnarray}
with $0 \le x \le 1$. Taking the nucleon at rest $P^\mu = 
(M_N, \mbox{\bf 0})$, we have $P^{+} = M_N / \sqrt{2}$ and then
\begin{eqnarray}
   q (x) \ \ = \ \ \frac{1}{4 \pi} \int^{\infty}_{-\infty} 
   d z^0 \,e^{-i x M_N z^0} \,\, 
   \langle N (\mbox{\boldmath $P$}) \,| \,\psi^\dagger (z) \,\gamma^0 \,
   \Gamma \,\psi(0) \,| \,N (\mbox{\boldmath $P$})\rangle \,\,
   |_{z^3 = -z^0,\, z_{\perp} = 0} \,\, . \ \ 
\end{eqnarray}
The basis of our calculation is the following path integral
representation of a matrix element of an arbitrary nonlocal quark
bilinear operator between nucleon states with definite momenta
[22] :
\begin{eqnarray}
   &\,& \langle N (\mbox{\boldmath $P$}) \,| \,\psi^\dagger (z) \,O \,
   \psi(0) \,| \,N (\mbox{\boldmath $P$})\rangle
   \ \ = \ \ \frac{1}{Z} \,\,\int \,\,d^3 x  \,\,d^3 y \,\,
   e^{\,- \,i \mbox{\boldmath $P$} \cdot \mbox{\boldmath $x$}} \,\,
   e^{\,i \,\mbox{\boldmath $P$} \cdot \mbox{\boldmath $y$}} \,\,
   \int {\cal D} U \nonumber \\
   &\times& \!\!\!\! \int {\cal D} \,\,
   \psi \,\,{\cal D} \,\,\psi^\dagger \,\,
   J_N (\frac{T}{2}, \mbox{\boldmath $x$}) \,\,\psi^\dagger (z) \,\,
   O \,\,\psi(0) \,\,J_N^\dagger (-\frac{T}{2}, \mbox{\boldmath $y$}) \,\,
   \exp \,[\,\,i \int \,d^4 x \,\,\bar{\psi} \,\,
   (\,i \! \not\!\partial \,- \,M
   U^{\gamma_5}) \,\psi \, ] \, , \ \ \ \   
\end{eqnarray}
where
\begin{eqnarray}
   {\cal L} \ \ = \ \ \bar{\psi} \,(\,i \not\!\partial \ - \ 
   M U^{\gamma_5} (x) \,) \,\psi \,\, ,
\end{eqnarray}
with $U^{\gamma_5} (x) = \exp [ \,i \gamma_5 \mbox{\boldmath $\tau$}
\cdot \mbox{\boldmath $\pi$} (x) / f_\pi \,]$ is the basic lagrangian
of the CQSM, and
\begin{equation}
   J_N (x) \ \ = \ \ \frac{1}{N_c !} \,\, 
   \epsilon^{\alpha_1 \cdots \alpha_{N_c}} \,\,
   \Gamma_{J J_3, T T_3}^{\{f_1 \cdots f_{N_c}\}} \,\,
   \psi_{\alpha_1 f_1} (x) \cdots \psi_{\alpha_{N_c} f_{N_c}} (x) \,\, ,
\end{equation}
is a composite operator carrying the quantum numbers $J J_3, T T_3$
(spin, isospin) of the nucleon, where $\alpha_i$ is the color index,
while $\Gamma^{\{ f_1 \cdots f_{N_C} \}}_{J J_3,T T_3}$ is a symmetric
matrix in spin-flavor indices $f_i$. By starting with a stationary
pion field configuration of hedgehog shape $U^{\gamma_5}_0
(\mbox{\boldmath $x$}) = \exp \,[ \,i \gamma_5 \mbox{\boldmath $\tau$}
\cdot \hat{\mbox{\boldmath $r$}} F(r) \,]$, the path integral over
the pion fields $U$ can be done in a saddle point approximation. Next, we
consider two important fluctuations around the static configuration,
i.e. the translational and rotational zero-modes.
To treat the translational zero-modes, we use an approximate
momentum projection procedure of the nucleon state, which amounts to
integrating over all shift $\mbox{\boldmath $R$}$ of the soliton
center-of-mass coordinates [24,19] :
\begin{equation}
   \langle N (\mbox{\boldmath $P$}^\prime ) \,| \,
   \psi^\dagger (z) \,O \,\psi (0) \,| \,N
   (\mbox{\boldmath $P$}) \rangle
   \ \longrightarrow \ \int \,\,d^3 R \,\,\,\,
   \langle N (\mbox{\boldmath $P$}^\prime ) \,| \,
   \psi^\dagger (z_0,\mbox{\boldmath $z$} - \mbox{\boldmath $R$}) \,
   O \,\psi (0,- \mbox{\boldmath $R$}) \,| \,N
   (\mbox{\boldmath $P$}) \rangle \,\, .
\end{equation}
The rotational zero-modes can be treated by introducing a rotating
meson field of the form :
\begin{eqnarray}
   U^{\gamma_5} ( \mbox{\boldmath $x$}, t) \ \ = \ \ A(t) \,\,
   U_0^{\gamma_5} (\mbox{\boldmath $x$}) \,\,A^\dagger (t) \,\, ,
\end{eqnarray}
where $A(t)$ is a time-dependent $SU(2)$ matrix in the isospin space.
Note first the identity
\begin{equation}
   \bar{\psi} \,( \,i \not\!\partial - M A (t) U^{\gamma_5}_0 
   (\mbox{\boldmath $x$}) A^\dagger (t) \,) \,\psi \ \ = \ \ 
   \psi^\dagger_A \,( i \partial_t - H - \Omega \,) \,\psi_A \,\, 
\end{equation}
with
\begin{eqnarray}
   \psi_A \ = \ A^\dagger (t) \,\psi \,\, , \hspace{5mm} 
   H \ = \ \frac{\mbox{\boldmath $\alpha$} \cdot \nabla}{i} 
   \ + \ M \,\beta \,U^{\gamma_5}_0 (\mbox{\boldmath $x$})
   \,\, , \hspace{5mm}
   \Omega \ = \ - \,i \,A^\dagger (t) \,\dot{A} (t) \,\, .
\end{eqnarray}
Here $H$ is a static Dirac Hamiltonian with the background pion
fields $U_0^{\gamma_5} (\mbox{\boldmath $x$})$, playing the role of a mean
field for quarks, while
$\Omega = \frac{1}{2} \Omega_a \tau_a$ is the SU(2)-valued
angular velocity matrix later to be quantized as $\Omega_a \rightarrow
\hat{J}_a / I$ with $I$ the moment of inertia of the soliton and
$\hat{J}_a$ the angular momentum operator.  
We then introduce a change of quark field variables $\psi \rightarrow
\psi_A$, which amounts to getting on a body-fixed rotating frame.
Denoting $\psi_A$ anew as $\psi$ for notational simplicity, the
nucleon matrix element (6) can then be written as
\begin{eqnarray}
   &\,& \langle N (\mbox{\boldmath $P$}) \,| \,\psi^\dagger (z) \,O \,
   \psi(0) \,   | \,N (\mbox{\boldmath $P$}) \rangle \nonumber \\
   &=& \frac{1}{Z} \,\,\Gamma^{\{ f \}} \,\,{\Gamma^{\{ g \}}}^* \,\,
   \int \,\,d^3 x \,\,d^3 y \,\,
   e^{-i \mbox{\boldmath $P$} \cdot \mbox{\boldmath $x$}} \,\,
   e^{i \mbox{\boldmath $P$} \cdot \mbox{\boldmath $y$}} \,\,
   \int \,\,d^3 R \nonumber \\
   &\times& \int \,\,{\cal D} A \,\,{\cal D} \psi \,\,
   {\cal D} \psi^\dagger
   \,\,\exp \,[ \,\,i \,\int \,d^4 x \,\,\psi^\dagger 
   (\,i \partial_t - H - \Omega) \,\psi \,] \nonumber \\
   &\times& \! \prod^{N_c}_{i = 1} \,\,\,
   [ \,A ( \frac{T}{2}) \,\,\psi_{f_i}
   ( \frac{T}{2}, \mbox{\boldmath $x$} ) \,] \,\,
   \psi^\dagger (z_0, \mbox{\boldmath $z$} - \mbox{\boldmath $R$})
   \,\,\tilde{O} \,\,
   \psi(0,- \,\mbox{\boldmath $R$}) \,\,
   \prod^{N_c}_{j = 1} \,\,\, [ \,\psi_{g_j}^\dagger
   (- \frac{T}{2}, \mbox{\boldmath $y$}) \,\,
   A^\dagger (-\frac{T}{2})] \,\, , \ \ 
\end{eqnarray}
with the definition $\tilde{O} \equiv A^\dagger (z_0) O A(0)$.
Now performing the path integral over the quark fields, we obtain
\begin{eqnarray}
   &\,& \langle N (\mbox{\boldmath $P$}) \,
   | \,\psi^\dagger (z) \,O \,\psi (0) \,
   | \,N (\mbox{\boldmath $P$}) \rangle  \nonumber \\
   &=& \frac{1}{Z} \,\,\tilde{\Gamma}^{\{ f \}} \,\,
   {\tilde{\Gamma}}^{{\{ g \}}^\dagger} \,\,N_c \,\,
   \int \,\,d^3 x \,\,d^3 y \,\,e^{\,-i \,\mbox{\boldmath $P$}
   \cdot \mbox{\boldmath $x$}}
   \,\,e^{\,i \mbox{\boldmath $P$} \cdot \mbox{\boldmath $y$}} \,\,
   \int d^3 R \,\,\int {\cal D} A \nonumber \\
   &\times& \{ \, {}_{f_1} \langle\frac{T}{2}, \mbox{\boldmath $x$}
   \,| \,\frac{i}
   {\,i \partial_t - H - \Omega} \,| \,z_0, \mbox{\boldmath $z$} - 
   \mbox{\boldmath $R$} \rangle_{\gamma} \cdot
   (\tilde{O})_{\gamma \delta} \cdot
   {}_\delta \langle 0, -\mbox{\boldmath $R$} \,| \,\frac{i}
   {\,i \partial_t - H - \Omega} \,
   | - \frac{T}{2}, \mbox{\boldmath $y$}\rangle_{g_1} \nonumber \\
   &-& \ \ \mbox{Tr} \,\,{( \,\langle0, -\mbox{\boldmath $R$} \,| \,
   \frac{i}{i \partial_t - H - \Omega} \,| \,
   z_0, \mbox{\boldmath $z$} - \mbox{\boldmath $R$}\rangle 
   \tilde{O} \,)} \,\,\,
   {}_{f_1} \langle \frac{T}{2},\mbox{\boldmath $x$} \,| \,
   \frac{i}{\,i \partial_t - H - \Omega} \,| - \frac{T}{2},
   \mbox{\boldmath $y$}\rangle_{g_1} \,\, \} \nonumber \\
   &\times& \prod^{N_c}_{j = 2} \,\,\,[ \,{}_{f_j}
   \langle \frac{T}{2}, \mbox{\boldmath $x$}
   \,| \,\frac{i}{\,i \partial_t - H - \Omega} \,
   | -\frac{T}{2}, \mbox{\boldmath $y$}\rangle_{g_j} \,] \,\cdot \,
   \exp \,[\,\mbox{Sp} \log \,(\,i \partial_t - H - \Omega) \, ] \,\, ,    
\end{eqnarray}
with $\tilde{\Gamma}^{\{f\}} = \Gamma^{\{f\}} \,
{[A(\frac{T}{2})]}^{N_c}$ etc. Here $\mbox{Tr}$ is to be taken over
spin-flavor indices. Assuming a slow rotation of the
hedgehog soliton, we can make use of an expansion in $\Omega$.
For an effective action, this gives
\begin{equation}
   \mbox{Sp} \log \,
   ( \,i \partial_t - H - \Omega) \ \ = \ \ 
   \mbox{Sp} \log \,(\,i \partial_t - H)
   \ + \ \,i \,\,\,\frac{1}{2} \,\,I \,\,
   \int \,\,\Omega_a^2 \,\,d t \,\, . 
\end{equation}
The second term here is essentially the action of a rigid rotor,
which plays the role of the evolution operator in the space of
collective coordinates. Using the expansion of the
single quark propagator as
\begin{eqnarray}
   &\,& {}_{f_1} \langle \frac{T}{2}, \mbox{\boldmath $x$} \,| \,
   \frac{i}{i \partial_t - H - \Omega} \,| \,z_0, 
   \mbox{\boldmath $z$} -    \mbox{\boldmath $R$} \rangle {}_\gamma
   \ \ = \ \ 
   {}_{f_1} \langle \frac{T}{2}, \mbox{\boldmath $x$} \,| \,
   \frac{i}{i \partial_t - H }
   \,| \,z_0, \mbox{\boldmath $z$} - 
   \mbox{\boldmath $R$} \rangle {}_\gamma \nonumber \\
   &\,& - \ \ \int \,\,d z_0^\prime \,\,
   d ^3 z^\prime \,\,{}_{f_1}
   \langle \frac{T}{2}, \mbox{\boldmath $x$} \,| \,
   \frac{i}{i \partial_t - H} \,
   | \,z_0^\prime, \mbox{\boldmath $z$}^\prime \rangle {}_\alpha \,\,
   \cdot \Omega_{\alpha \beta} (z_0^\prime) \cdot
   {}_\beta \langle z_0^\prime, \mbox{\boldmath $z$}^\prime \,| \,
   \frac{i}{i \partial_t - H} \,| \,z_0, \mbox{\boldmath $z$} - 
   \mbox{\boldmath $R$} \rangle {}_\gamma \nonumber \\
   &\,& + \hspace{65mm} \dots \,\,\,\, , 
\end{eqnarray}
we can separate the zeroth order ($\sim \! N^0_c$) and the first order
($\sim \! 1 / N_c$) corrections in $\Omega$ for nucleon observables.
Here we are interested in the linear order term in $\Omega$, because
the isovector unpolarized and isoscalar polarized distribution
functions both vanish at the lowest order. Since necessary
manipulations were explained in the previous papers [22,23],
we shall skip the detail except what seems absolutely necessary.
The only but important difference with the previous case is that
we are here handling a nucleon matrix element of a
quark bilinear operator which is {\it nonlocal \/} both in space
and time coordinates. This is only natural,
because we are investigating here a
quark-quark correlation function with a light-cone separation.
First by neglecting the time-ordering of two operators
$\Omega$ and $\tilde{O}$, the $O (\Omega^1)$ contribution to
(14) would become
\begin{eqnarray}
   &\,& \langle N (\mbox{\boldmath $P$}) \,| \,\psi^\dagger (z) \,
   O \,\psi (0) \,| \,
   N (\mbox{\boldmath $P$}) \rangle^{\Omega^1}  \nonumber \\
   &=& \frac{1}{Z} \,\,{\,\tilde{\Gamma}}^{\{ f \}} \,\,
   {\tilde{\Gamma}}^{{\{ g \}}^\dagger} \,\,N_c \,\, 
   \int \,\,d^3 x \,\,d^3 y  \,\,e^{\,-i \mbox{\boldmath $P$} 
   \cdot \mbox{\boldmath $x$}} 
   \,\,e^{\,i \mbox{\boldmath $P$} \cdot \mbox{\boldmath $y$}}
   \,\,\int \,\,d^3 R \,\,
   \int d^3 z^{\prime} \,\,d z_0^\prime \,\,\,\,
   \Omega_{\alpha \beta} (z_0^\prime) \,\,
   \tilde{O}_{\gamma \delta} (z_0, 0) \nonumber \\
   &\times& \!\!\!\! \{ \,{}_{f_1}
   \langle\frac{T}{2}, \mbox{\boldmath $x$} \,| \, 
   \frac{i}{i \partial_t - H} \,| \,
   z_0^\prime, \mbox{\boldmath $z$}^\prime \rangle_{\alpha} \cdot
   {}_ {\beta} \langle z_0^\prime, \mbox{\boldmath $z$}^\prime 
   \,| \,\frac{i}{i \partial_t - H} \,|\, z_0, \mbox{\boldmath $z$} 
   - \mbox{\boldmath $R$}\rangle_\gamma  \cdot {}_\delta \langle 0, - 
   \mbox{\boldmath $R$}
   \,| \,\frac{i}{i \partial_t - H} \,| \,- 
   \frac{T}{2}, \mbox{\boldmath $y$}\rangle_{g_1} \ \ \ \ \nonumber \\
   &+& \!\!\!\! {}_{f_1} \langle \frac{T}{2}, 
   \mbox{\boldmath $x$} \,| \,
   \frac{i}{i \partial_t - H} \,| \,
   z_0, \mbox{\boldmath $z$} - \mbox{\boldmath $R$} \rangle_{\gamma}
   \cdot {}_{\delta} \langle0 - \mbox{\boldmath $R$} \,| \,
   \frac{i}{i \partial_t - H} \,| \,z_0^\prime, 
   \mbox{\boldmath $z$}^\prime \rangle_{\alpha} \cdot {}_{\beta}
   \langle z_0^\prime, 
   \mbox{\boldmath $z$}^\prime \,| \,\frac{i}{i \partial_t -H} | 
   - \frac{T}{2},
   \mbox{\boldmath $y$}\rangle_{g_1} \ \ \ \ \nonumber \\
   &-& \!\!\!\! {}_{f_1} \langle \frac{T}{2}, \mbox{\boldmath $x$}
   \,|\, \frac{i}{i \partial_t - H} \,| \,
   - \frac{T}{2}, \mbox{\boldmath $y$} \rangle_{g_1} \cdot
   {}_{\delta} \langle 0, -\mbox{\boldmath $R$} \,| 
   \,\frac{i}{i \partial_t - H}
   \, | \,  z_0^\prime, \mbox{\boldmath $z$}^\prime\rangle_{\alpha}
   \cdot {}_{\beta} \langle z_0^\prime, 
   \mbox{\boldmath $z$}^\prime \,| \, 
   \frac{i}{i \partial_t - H} \,| \, z_0, \mbox{\boldmath $z$} - 
   \mbox{\boldmath $R$}\rangle_{\gamma} \} \ \ \ \ \nonumber \\
   &\times& \prod^{N_c}_{j = 2} \,\,\,[\, {}_{f_j}
   \langle \frac{T}{2}, \mbox{\boldmath $x$}
   \,| \,\frac{i}{i \partial_t - H} \,| \,
   - \frac{T}{2}, \mbox{\boldmath $y$} \rangle_{g_j} \,] \,\cdot \,
   \,\,\,\exp \,[\,\,\mbox{Sp} \log (\,i \partial_t - H)
   \,+ \,i \,\frac{T}{2} 
   \,\int \Omega_a^2 \,d t \,\,]  \,\, ,
\end{eqnarray}
As was extensively argued in [22] and [23], one must be careful about
the time order of two operators $\Omega$ and $\tilde{O}$, which
do not generally commute after collective quantization of the
rotational zero-energy modes. In the present case, the proper
account of this time-odering can be achieved by the
following replacement in the equation above :
\begin{eqnarray}
   \Omega_{\alpha \beta} (z_0^\prime) \,\,\tilde{O}_{\gamma \delta}
   (z_0, 0)  &\longrightarrow& \,
   [ \,\,\theta( z_0^\prime, z_0, 0) 
   \,+ \,\theta( z_0^\prime, 0, z_0) \,\,] \,\,\Omega_{\alpha \beta} \,\, 
   \tilde{O}_{\gamma \delta} \nonumber \\
   &+& \,
   [ \,\,\theta( z_0, 0, z_0^\prime)
   \,+ \,\theta(0, z_0, z_0^\prime) \,\,] \,\,\tilde{O}_{\gamma \delta} 
   \,\,\Omega_{\alpha \beta} \,\, ,
\end{eqnarray}
where $\theta (a,b,c)$ is a short-hand notation of a step function
which is $1$ when $a > b > c$ and $0$ otherwise. 
Here we have intentionally dropped terms containing a factor
$\theta (z_0,z'_0,0)$ or $\theta (0,z'_0,z_0)$.
(Note that, if $z_0^\prime$ is allowed to lie between $z_0$ and
$0$, the above idea of the time order of the two operators 
$\Omega$ and $\tilde{O}$ would lose its definite meaning.)
It amounts to excluding diagrams in which the
Coriolis force $\Omega$ operates in the time interval between $z_0$
and $0$. This is motivated by the physical picture that a deep
inelastic scattering process is a short distance phenomenon and
its typical time scale is much shorter than that of the collective
rotational motion which we assume is much slower than the velocity of
the intrinsic quark motion. Then, one should rather treat $z_0$ and $0$
as nearly degenerate when performing the expansion in $\Omega$, which
dictates the neglect of diagrams containing
$\theta (z_0,z'_0,0)$ or $\theta (0,z'_0,z_0)$.
The exclusion of these special
time-order diagrams has an important physical consequence as we
shall come back later. Another important remark is that we do not
give any constraint on the time
order of $z_0$ and $0$ themselves : $z_0$ can be earlier or later
than $0$. It can be verified that inclusion of both possibilities is
essential for maintaining the charge conjugation symmetry,
which denotes that the quark (antiquark) distributions in the
nucleon must be the same as the antiquark (quark) distributions
in the antinucleon.
The last comment we want to make here is on the treatment of the
rotated operator $\tilde{O} = A^\dagger (z_0) \,\gamma_0 \,O \,A(0)$.
If the operator contains an isospin factor (or it is an isovector
operator), we are to use the identity
\begin{equation}
   A^{\dagger} (z_0) \tau_a A(0) 
   \ \ = \ \ \frac{1}{2} \,\,\mbox{Tr} 
   (A^{\dagger} (z_0) \,\tau_a \,A (0) \,\tau_b) \,\,\tau_b \,\, .
\end{equation}
A new feature here is that the factor
$\mbox{Tr} (A^\dagger (z_0) \,\tau_a \,A(0) \,\tau_b)$
contains two time arguments $z_0$ and $0$. Still, we
can use the usual procedure in which it
is replaced by the standard Wigner function $D_{ab} (A)$.
Intuitively, the justification follows from our dynamical
assumption that $z_0$ and $0$ can
be regarded as nearly degenerate as compared with the time scale of
the collective rotational motion. A formally more rigorous
justification for this replacement may be obtained by remembering
that the time evolution of the collective space
operator $A$ is given by
\begin{equation}
   A(z_0) \ = \ \exp( \,i \,z_0 \,\mbox{\boldmath $J$}^2 / I \,) \,\,
   A(0) \,\, \exp( \,- \,i \,z_0 \,\mbox{\boldmath $J$}^2 / I \,) \,\, ,
\end{equation}
with $\mbox{\boldmath $J$}$
the angular momentum operator working in the collective coordinate
space and $I$ the moment of inertia of the soliton.
Since the moment of inertia $I$ is an $O (N_c)$ quantity, this gives
\begin{equation}
   \frac{1}{2} \,\,\mbox{Tr} \,\,( \,A^\dagger (z_0) \,\,\tau_a \,\,
   A(0) \,\,\tau_b \,) \ \ = \ \ D_{ab} (A) \,\,[ \, 1 \ + \ 
   O (1 / N_c) \,] \,\, ,
\end{equation}
convincing that the error of the above replacement is a higher
order correction in the $1 / N_c$ expansion.

After stating all the delicacies inherent in the structure function
problem, we can now proceed in the same way as [22] and [23]. Using
the spectral representation of the single quark Green function
\begin{eqnarray}
   {}_{\alpha} \langle\mbox{\boldmath $x$}, t \,|\,
   \frac{i}{i \partial_t - H} \,| \,
   \mbox{\boldmath $x$}^\prime, t^\prime\rangle_{\beta}
   &=& \theta(t - t^\prime) \,\,\sum_{m>0} \,\,
   e^{\,- i E_m (t - t^\prime)} \,\,
   {}_{\alpha} \langle\mbox{\boldmath $x$}\,| \,m \rangle 
   \langle m \,| \,\mbox{\boldmath $x$}^\prime \rangle_{\beta}
   \nonumber \\
   &-& \theta(t^\prime - t) \,\,\sum_{m<0} \,\, 
   e^{\,-i E_m (t - t^\prime)} \,\,
   {}_{\alpha} \langle\mbox{\boldmath $x$} \,| \,m \rangle
   \langle m \,| \,\mbox{\boldmath $x$}^\prime\rangle_{\beta} \,\, , \ \  
\end{eqnarray}
together with the relation
\begin{equation}
   \langle \mbox{\boldmath $z$} - \mbox{\boldmath $R$} \, | \, 
   \ \ = \ \ \langle -\mbox{\boldmath $R$} \,| \,\,
   e^{\,i \mbox{\boldmath $p$} \cdot \mbox{\boldmath $z$}} \,\, 
\end{equation}
we can perform the integration over $\mbox{\boldmath $R$}, 
\mbox{\boldmath $z$}^\prime$, and $z'_0$.
The resultant expression is then put into (5) to carry out the
integration over $z_0$. We then arrive at the formula, which gives
the theoretical basis for evaluating the first order
contributions in $\Omega$ to quark distribution functions of the
nucleon :
\begin{equation}
   q (x ; \Omega^1) \ \ = \ \ \int \,\,d A \,\,\,
   \Psi_{J_3 T_3}^{{(J)}^*} [A] \,\,\,O^{(1)} [A] \,\,\,
   \Psi_{J_3 T_3}^{(J)} [A] \,\,\, .
\end{equation}
Here
\begin{equation}
   \Psi_{J_3 T_3}^{(J)} \ \ = \ \ \sqrt{\frac{2J+1}{8 \pi^2}} \,\,
   {(-1)}^{T + T_3} \,\,D_{-T_3 J_3}^{(J)} (A) \,\, ,
\end{equation}
are wave functions, describing the collective rotational motion
of the hedgehog soliton, while
\begin{equation}
   O^{(1)} [A] \ \ = \ \ O_{val}^{(1)} \,[A] \ + \ 
   O_{v.p.}^{(1)} \,[A] \,\, ,
\end{equation}
where
\begin{eqnarray}
   &\,& O_{val}^{(1)} \ \ = \ \ M_N \cdot N_c \nonumber \\
   &\times& \!\! \{ \,\,\sum_{m>0} \,\frac{1}{E_m - E_0} \,
   \langle 0 \,| \,\tilde{O} \,\frac{\delta(x M_N - E_0 - p^3) +
   \delta(x M_N - E_m - p^3)}{2} \,| \,m \rangle \,\,
   \langle m \,| \,\Omega \,| \,0 \rangle  \ \ \ \ \ \ \nonumber \\
   &+& \,\,\sum_{m<0} \,\frac{1}{E_m - E_0} \,
   \langle m \,| \,\tilde{O} \,\frac{\delta(x M_N - E_0 - p^3) +
   \delta(x M_N - E_m - p^3)}{2} \,| \,0 \rangle \,\,
   \langle 0 \,| \, \Omega \,| \,m \rangle \, \}
   \ \ \ \ \ \ \nonumber \\
   &+& \hspace{60mm} ( \, \tilde{O} \leftrightarrow \Omega \, )
   \,\,\,\,\, . \hspace{60mm}
\end{eqnarray}
and
\begin{eqnarray}
   &\,& O_{v.p.}^{(1)} \ \ = \ \ M_N \cdot N_c \,\,
   \sum_{m \ge 0, n < 0} \,\frac{1}{E_m - E_n} \nonumber \\
   &\times& \!\!\!\! \{ \,< n \,| \,\,\tilde{O} \,\,
   \frac{\delta(x M_N - E_n - p^3) +
   \delta(x M_N - E_m - p^3)}{2} \,| \,m \rangle \,\,
   \langle m \,| \,\Omega \,| \,n \rangle \,+ \, 
   (\tilde{O} \leftrightarrow \Omega) \,\} \, , \ \ \ \ \ \ 
\end{eqnarray}
with $p^3$ the $z$-component of the single-quark momentum operator.
In the above formulas, $| \,m \rangle$ and $E_m$ denote the eigenstates
and associated eigenenergies of the static Dirac Hamiltonian $H$.
In particular, $| \,0 \rangle$ represents the lowest energy eigenstate
of $H$, which emerges from the positive energy Dirac continuum.
This particular state is referred to as the valence quark orbital
following the convention of previous studies [16-17]. However,
the term ``valence quark'' here should not be confused with the
corresponding term in the language of the quark parton model.  
The two contributions $O^{(1)}_{val}$ and $O^{(1)}_{v.p.}$ can
actually be put together into the form :
\begin{eqnarray}
   &\,& O^{(1)} \ \ = \ \ M_N \cdot N_c \,\,
   \sum_{m > 0, n \le 0} \,\frac{1}{E_m - E_n} \nonumber \\
   &\times& \!\!\!\! \{ \,\langle n \,| \,\tilde{O} \,
   \frac{\delta(x M_N - E_n - p^3) +
   \delta(x M_N - E_m - p^3)}{2} \,| \,m \rangle \,\,
   \langle m \,| \,\Omega \,| \,n \rangle \,+ \,
   (\tilde{O} \leftrightarrow \Omega) \,\} \, , \ \ \ \ \ \ 
\end{eqnarray}
which contains transitions from occupied single quark levels
to unoccupied ones in consistent with the Pauli principle [23].
It is important to recognize that this
nice feature would not follow, if one would include the
afore-mentioned particular time-order diagrams containing
$\theta (z_0,z'_0,0)$ and $\theta (0,z'_0,z_0)$.
In fact, if we had included these contributions by forgetting about
the time order of $\Omega$ and $\tilde{O}$, we would have
obtained
\begin{eqnarray}
   &\,& O_{v.p.}^{(1)} \ \ = \ \ M_N \cdot \frac{N_c}{2} \,\,
   \sum_{m, n} \,\frac{1}{E_m - E_n} \nonumber \\
   &\times& \!\!\!\!\! \langle n \,| \,\tilde{O} \,
   \frac{\mbox{sign} (E_m) \,\delta(x M_N - E_m - p^3) -
   \mbox{sign} (E_n) \,\delta(x M_N - E_n - p^3)}{2} \,| \,m \rangle \,
   \langle m \,| \,\Omega \,| \,n \rangle \, , \ \ \ \ \ 
\end{eqnarray}
instead of (28). This essentially coincides with the equation
which is obtained from (4.12) in ref.[19]. Clearly, the
formula (30) contains unpleasant Pauli-violating contributions
due to transitions between occupied levels themselves and
unoccupied ones. It is very important to recognize that this
Pauli-violating contributions to a quark distribution function
has nothing to do with the regularization problem. (This is
clear, since we have not yet introduced any regularization at
this stage.) This kind of Pauli-principle violation has
never been observed when only the nucleon matrix elements of
local quark bilinear operators have been dealt with.
In fact, one can confirm it very easily, if one carries out
the $x$ integration of (30) from $x = -1$ to $x = 1$, which
should give the first moment of the relevant distribution
function being expressed as a nucleon matrix element of
a local quark bilinear operator. One certainly convinces that
terms with $E_m \times E_n > 0$ drop completely.

We also call attention to the fact that the delta-function
representing the on-shell energy conservation appears in the
symmetrized form with respect to the energies of the single-quark
levels $| \,m \rangle$ and $| \,n \rangle$. The appearance of these two
delta-functions can be traced back to two possible time
order of $z_0$ and $0$ in $\bar{\psi} (z) \Gamma \psi (0)$,
and it plays a favorable role to maintain the charge-conjugation
symmetry as already mentioned.
We point out that a naive cranking procedure without taking 
care of the non-local (in time space) nature of the quark
bilinear operator would violate this property [18].

Using the general formulae given in (24) $\sim$ (28),
we can readily obtain the following expression for the isovector
unpolarized distribution function for the proton :
\begin{eqnarray}
   &\,& u(x) - d(x) \ = \ M_N \cdot \frac{1}{I} \cdot \frac{1}{3} \,\,
   \sum_{a=1}^3 \,\,\,\frac{N_c}{2} \,\,\sum_{m > 0, n \le 0} \,\,
   \frac{1}{E_m - E_n} \nonumber \\
   &\times& \!\! \langle n \,| \,\,\tau_a \,
   ( \,1 \,+ \,\gamma^0 \,\gamma^3 \,) \,\,
   \frac{\delta(x M_N - E_n - p^3) +
   \delta(x M_N - E_m - p^3)}{2} \,| \,m \rangle \,\,
   \langle m \,| \,\tau_3 \,| \,n \rangle  \,\, . \ \ \ \ \ \ \ 
\end{eqnarray}
(In the actual numerical calculation, we use the expression
in which the contribution of the valence quark level is separated
from that of the Dirac continuum according to the general
formulas (26) $\sim$ (28), since only the vacuum polarization part
is to be regularized in our regularization scheme.)

By integrating $u(x) - d(x)$ over $x$ between $-1$ and $1$,
the isospin sum rule follows;
\begin{equation}
   \int_{-1}^1 \,\,[ \,u(x) - d(x) \,] \,\, dx \ = \ 
   \int_0^1 \,\, \{ \,[u(x) - d(x)] - 
   [ \bar{u} (x) - \bar{d} (x) \,] \,\} \,\, dx \ = \ 1 \,\, ,
\end{equation}
provided that the moment of inertia is given by
\begin{equation}
   I \ \ = \ \ \frac{N_c}{2} \,\,
   \sum_{m > 0, n \le 0} \,\, \frac{1}{E_m - E_n} \,\,
   \langle n \,| \,\tau_3 \,| \,m \rangle \,\,
   \langle m \,| \,\tau_3 \,| \,n \rangle \,\, .
\end{equation}
On the other hand, the Gottfried sum is given as 
\begin{eqnarray}
   S_G \ &=& \ \frac{1}{3} \,\,\int_0^1 \,\,
   [\, u(x) - d(x) + \bar{u} (x) - \bar{d} (x) \,]
   \,\,dx \nonumber \\
   &=& \ \frac{1}{3} \ \ + \ \ \frac{2}{3} \,\,\int_0^1 \,\,
   [ \,\bar{u} (x) - \bar{d} (x) \,] \,\, dx \,\, .
\end{eqnarray}

The formula for the isosinglet polarized distribution 
function can similarly be written down as
\begin{eqnarray}
   &\,& \Delta u(x) + \Delta d(x) \ \ = \ \ - \,\,M_N \cdot
   \frac{1}{I} \cdot \frac{N_c}{2} \,\,\sum_{m > 0, n \le 0} \,\,
   \frac{1}{E_m - E_n} \nonumber \\
   &\times& \!\!\! \langle n \,| \,(1 + \gamma^0 \gamma^3 ) \,\gamma_5 \,\,
   \frac{\delta(x M_N - E_n - p^3) + \delta(x M_N - E_m - p^3)}{2}
   \,| \,m \rangle \,\langle m \,| \,\tau_3 \,| \,n \rangle \,\, . \ \ \ \ \ 
\end{eqnarray}
Integrating over $x$, we get
\begin{eqnarray}
   \int_{-1}^1 \,\, [ \, \Delta u(x) \!\!\! &+& \!\!\! \Delta d(x) \,]
   \,\,\, dx 
   \ = \ \int_0^1 \,\,[ \,\Delta u(x) + \Delta d(x) + 
   \Delta \bar{u} (x) + \Delta \bar{d} (x) \,] \,\,\, dx \nonumber \\
   &=& - \,\,\frac{1}{I} \cdot \frac{N_c}{2} \,\,
   \sum_{m > 0, n \le 0} \,\,\frac{1}{E_m - E_n} \,\,
   \langle n \,| \,\gamma^0 \,\gamma^3 \,\gamma_5 \,| m \rangle \,
   \langle m \,| \,\tau_3 \,| \, n \rangle \,\, ,
\end{eqnarray}
which is noting but the isosinglet axial-vector charge ( or the quark
spin fraction) of the nucleon first investigated
in [16] in the context of the CQSM.

\vspace{4mm}
\section{Numerical results and discussion}

\ \ \ Now we turn to the discussion on the method of numerical
calculation. For evaluating isovector unpolarized and isoscalar
polarized distribution functions with inclusion of the Dirac-sea
quark degrees of freedom, we must perform infinite double sums over
all the single-quark orbitals which are eigenstates of the
static hamiltonian $H$. A numerical technique for carrying out
such double sums has been established in the case of static nucleon
observables [16]. In the nucleon structure function problem, however,
we encounter a new bothering feature. The relevant matrix
elements contain delta functions depending on the single-quark
momentum operator $p^3$. Kahana and Ripka's plane wave basis [25]
is useful also for treating this feature.
(We recall that it is a set of eigenstates of
the free hamiltonian $H_0 = \frac{\mbox{\boldmath $\alpha$} 
\cdot \nabla}{i} + \beta M$ discretized by imposing an appropriate
boundary condition for the radial wave functions at the radius $D$
chosen to be sufficiently larger than the soliton size.)
For evaluating necessary matrix elements, we find it convenient to
expand the eigenstates $| \,m \rangle$ and
$| \,n \rangle$ of $H$ in terms of this discretized plane wave basis
in the momentum representation. We can then make use of the fact
that $p^3$ is diagonal in this
basis. This however does not get rid of all the trouble.
Since the delta function in question now depends on the Bjorken
variable $x$, $E_m$ (or $E_n$), $p^3 = | \mbox{\boldmath $p$} |_i \,
\cos \theta_p$, where $| \mbox{\boldmath $p$} |_i$ are the momenta of
the discretized basis states, the resultant distribution would be
a discontinuous function of $x$ [20].
To remedy it, Diakonov et al. proposed a smearing procedure
in which a calculated distribution function is convoluted with a
Gaussian distribution with an appropriate width [20].
This smearing procedure works well as a whole, but it turns out
to have some unpleasant features. First, it requires us a special
care near $x \simeq 0$, where a distribution function (extended to
the range $-1 \le x \le 1$) or its derivative may have a
discontinuity. Secondly, the smearing causes unpleasant flattening
when the bare distribution has a sharp peak as is the case for
the valence quark contribution. Then, we decided to
use here a conceptually simpler least square fitting
procedure. To this end, we first calculate distribution functions
at many points of $x$ with a small interval,
say $\Delta x = 0.01$.
This gives a set of data, which are rapidly
fluctuating with a small amplitude. A common variance of the order
of this fluctuation amplitude is then given to all the data points,
so that the least-square fit can be carried out.
In the fitting of the distribution functions, we find it convenient
to use a set of locally peaked Gaussian functions, the centers of
which are distributed homogeneously in the $x$ space.

Actually, the expression for the isovector unpolarized
distribution as well as that for the moment of inertia contain
divergences, so that they need some regularization. In the case of
static nucleon observables, there is a phenomenologically
established regularization method, based on Schwinger's
proper-time scheme [16,17]. Unfortunately, how to generalize this
regularization scheme in the evaluation of nucleon structure
functions has not been known yet.
In the recent investigations of the leading order quark
distribution functions [19,20], Diakonov et al. advocated to
use the Pauli-Villars regularization.
In this scheme, one subtracts from divergent sums a multiple
of the corresponding sums over eigenstates of the hamiltonian
in which the quark mass $M$ is replaced by the regulator mass
$M_{PV}$, uniquely determined from the condition,
\begin{equation}
   f^2_\pi \ \ = \ \ \frac{N_c \,M^2}{4 \,\pi^2} \,\,\ln \,\,
   \frac{M^2_{PV}}{M^2} \,\, ,
\end{equation}
which gives $M_{PV} \simeq 571 \,\mbox{MeV}$ for $M = 400 \,
\mbox{MeV}$. 
In the present study, we shall also try this regularization scheme
but somewhat in a different way. That is, although they regularized
not only the Dirac continuum contribution but also the discrete
(valence) level one, it is not an absolute demand of this
regularization scheme. Since the valence level contribution is
convergent itself, one can also define and use another regularization
scheme, in which only the Dirac sea part is regularized by the
Pauli-Villars subtraction. In fact, a closer correspondence with
the standardly-used regularization scheme like the proper-time one
is held by adopting this latter scheme of regularization.
Either way, since this method of regularization with use of
subtraction method is quite dissimilar to more standard
energy-cutoff scheme, one may suspect that the answers would depend
strongly on this particular choice of regularization.
To see how much the predicted distributions depend on the choice
of the regularization method, we have invented another
semi-theoretical regularization scheme, which we believe is
closer in spirit to the standard method of introducing regularization.
Let us explain it for $u(x) - d(x)$,
since the vacuum quark part of $\Delta u(x) + \Delta d(x)$ is
convergent without regularization. We start with the nonregularized
form for the vacuum polarization part of $u(x) - d(x)$
\begin{eqnarray}
   \!\!\!\!\! &\,& {[u(x) - d(x)]}_{v.p.} \ = \ 
   M_N \cdot \frac{1}{I} \cdot \frac{1}{3} \,\,
   \sum_{a=1}^3 \,\,\,\frac{N_c}{2} \,\,\sum_{m \ge 0, n < 0} \,\,
   \frac{1}{E_m - E_n} \hspace{30mm} \nonumber \\
   &\times& \!\!\! < n \,| \,\,\tau_a \,
   ( \,1 \,+ \,\gamma^0 \,\gamma^3 \,) \,\,
   \frac{\delta(x M_N - E_n - p^3) +
   \delta(x M_N - E_m - p^3)}{2} \,| \,m > \,\,
   < m \,| \,\tau_3 \,| \,n >  \,\, . \hspace{1cm}
\end{eqnarray}
One of the simplest choice for introducing a regularization, which
does not violate the charge-conjugation symmetry, would be
given by the replacement,
\begin{eqnarray}
   &\,& \frac{\delta(x M_N - E_n - p^3) + 
   \delta(x M_N - E_m - p^3)}{2} \nonumber \\
   &\longrightarrow& \ \ 
   \frac{g(E_n ; E_{max}) \,\delta(x M_N - E_n - p^3) + 
   g(E_m ; E_{max}) \,\delta(x M_N - E_m - p^3)}{2} 
   \,\, , \hspace{15mm}
\end{eqnarray}
where $g(E_n ; E_{max})$ is an appropriate energy cutoff which
may, for instance, be taken as
\begin{equation}
   g(E_n ; E_{max}) \ = \ \mbox{erfc} (|E_n| / E_{max}) \,\, ,
\end{equation}
or
\begin{equation}
   g(E_n ; E_{max}) \ = \ \exp( - E_n^2 / E_{max}^2 ) \,\, .
\end{equation}
For definiteness, we will take the second choice in
the following explanation.
Integrating over $x$ from $-1$ to $1$, this leads to the Dirac sea
contribution to the moment of inertia of the form :
\begin{equation}
   I_{v.p.} \ = \ \frac{N_c}{2} \,\sum_{m \ge 0, n < 0} \,
   \frac{1}{E_m - E_n} \cdot
   \frac{e^{-E_n^2 / E_{max}^2} + e^{-E_m^2 / E_{max}^2}}{2}
   \,\,
   \langle n \,| \,\tau_3 \,| \,m \rangle \,
   \langle m \,| \,\tau_3 \,| \,n \rangle ,
\end{equation}
where use has been made of the symmetry of the matrix element
of $\tau_3$.
One notices that this expression is pretty different from that
of the standardly-used proper-time regularization,
which gives 
\begin{equation}
   I_{v.p.} \ = \ \frac{N_c}{2} \,\sum_{m, n} \,
   f(E_m, E_n ; \Lambda) \,
   < n \,| \,\tau_3 \,| \,m > \,< m \,| \,\tau_3 \,| \,n > ,
\end{equation}
with
\begin{eqnarray}
   f(E_m, E_n : \Lambda) &=& \frac{1}{4} \,\, \cdot \{ \,
   \frac{\mbox{sign} (E_m) \,\mbox{erfc} (|E_m| / \Lambda) - 
   \mbox{sign} (E_n) \,\mbox{erfc} (|E_n| / \Lambda)}{E_m - E_n}
   \nonumber \\
   &\,& \hspace{1cm} - \,\,\frac{2}{\sqrt{\pi}} \,\,\Lambda \,\,
   \frac{e^{-E_m^2 / \Lambda^2} - 
   e^{-E_n^2 / \Lambda}}{E_m^2 - E_n^2} \,\} \,\, , 
\end{eqnarray}
though both are reduced to the following same expression in the
infinite cutoff limit :
\begin{equation}
   I_{v.p.} \ = \ \frac{N_c}{2} \,\sum_{m \ge 0, n < 0} \,
   \frac{1}{E_m - E_n} \,
   < n \,| \,\tau_3 \,| \,m > \,< m \,| \,\tau_3 \,| \,n >  \,\, .
\end{equation}
A noticeable difference is that the formulas based on the
proper-time regularization contains transitions between 
occupied levels themselves and unoccupied ones,
which clearly violates Pauli exclusion principle. However, this
physically unacceptable Pauli-principle violation of the
proper-time regularization is a mere appearance.
Due to the favorable function of the second term of the cutoff
function $f(E_m, E_n ; \Lambda)$, it turns out that the
contribution of the Pauli-violating processes with
$E_m \times E_n > 0$ turns out to be greatly suppressed.
On the other hand, the regularized expression (42) is free
from the Pauli-principle violation by construction. However, since
the behavior of the cutoff functions in (42) and (43) are
pretty different, there is no reason to expect that the
cutoff parameter $E_{max}$ in (41) is close to $\Lambda$ in (44).
That means that we need some criterion to determine the
parameter $E_{max}$ in (40) or (41).
In view of the fundamental importance of the moment of inertia
in the collective quantization approach, a natural procedure
would be to fix the cutoff parameter so as to reproduce
this key quantity. One may then fix $E_{max}$ so that the new
regularization scheme gives the same moment of inertia as
the standard proper-time scheme. Here, for the sake of convenience,
we would rather fix it so that the new regularization scheme
gives the same moment of inertia as the Pauli-Villars regularization
scheme. The reason is as follows. In the previous studies of
static nucleon observables, all the observables can be evaluated
within the single regularization scheme, which also enables us
to get a self-consistent soliton solution [16,17].
In the present semi-theoretical treatment of the regularization
problem, we must give up this complete self-consistency.
Under this circumstance, the best starting choice
would be to use a phenomenologically tested soliton profile,
which has been obtained in the proper-time regularization scheme
with use of the dynamical quark mass $M = 400 \,\mbox{MeV}$ and the
physical pion mass $m_\pi = 138 \,\mbox{MeV}$. Using this
soliton profile, we obtain $I_{total} = 1.16 \,\mbox{fm}$ in
the proper-time regularization scheme, while the Pauli-Villars
regularization scheme with the same profile function gives
$I_{total} = 1.14 \,\mbox{fm}$.
This means that the above two criterions for fixing $E_{max}$ can be
thought of as practically the same. Now by setting as
\begin{eqnarray}
   E_{max} &=& \ \,728 \,\mbox{MeV} \hspace{8mm} \mbox{for} 
   \hspace{8mm}
   g(E_n ; E_{max}) \ \ = \ \ \exp (-E_n^2 / E_{max}^2 ) \,\, ,
   \nonumber \\
   E_{max} &=& 1080 \,\mbox{MeV} \hspace{8mm} \mbox{for} \hspace{8mm}
   g(E_n ; E_{max}) \ \ = \ \ \mbox{erfc} (|E_n| / E_{max}) \,\, ,
\end{eqnarray}
the new regularization schemes give
\begin{eqnarray}
   I_{v.p.} (\mbox{Gaussian-function}) &\simeq& 0.254 \,\mbox{fm}
   \,\, , \nonumber \\
   I_{v.p.} (\mbox{error-function}) \ \ \ &\simeq& 0.254 \,\mbox{fm}
\end{eqnarray}
which approximately reproduce the number
\begin{equation}
   I_{v.p.} (\mbox{Pauli-Villars}) \ \ \simeq \ \ 0.255 \,\mbox{fm},
\end{equation}
obtained with the Pauli-Villars regularization scheme.
Thus, at least once an appropriate soliton profile is given,
either of these three regularization schemes thought expected
to be a reasonable candidate
for regularizing divergences in our problem. In the following,
we shall calculate the vacuum polarization contributions to
$u(x) - d(x)$ and $\bar{u} (x) - \bar{d} (x)$ by using these
three different schemes of regularization, to see the
regularization dependence of distribution functions,.

Now we are in a position to show the results of our numerical
calculation. Shown in Fig.1 are the contributions of the (discrete)
valence level to the distribution functions $u(x) - d(x)$ and
$\bar{u} (x) - \bar{d} (x)$ in the proton. One clearly sees
that $\bar{u} (x) - \bar{d} (x)$ is negative in the range of
$x$ where it has dominant support. This means that the excess
of the $\bar{d}$ sea over the $\bar{u}$ in the proton comes
out very naturally even at the level of valence quark
approximation. This feature, which was also observed in ref.[18],
is an interesting consequence of the CQSM, which assumes
the symmetry breaking mean-field of hedgehog form.
One may be interested in what conclusion one would
obtain if one evaluates the net flavor asymmetry of sea quarks
under this valence quark approximation. 
One gets
\begin{equation}
   \int_0^1 \,\,{[ \,\bar{u} (x) - \bar{d} (x) \,]}_{val}
   \,\, dx \ \ \simeq \ \ - \,0.017 \,\, .
\end{equation}
(Here the upper limit of $x$ integration is actually extended
to $\infty$, since the theoretical distribution
functions have non-zero support beyond $x=1$. See the discussion
before (52).) The above number may be compared with the
empirical estimate
\begin{equation}
   \int_0^1 \,\,[ \,\bar{u} (x) - \bar{d} (x) \,] \,\,dx 
   \ \ \simeq \ \ - \,0.148 \,\, ,
\end{equation}
which is obtained from the NMC measurement $S_G = 0.235 
\pm 0.026$ [26].
One sees that the valence quark contribution alone is not
enough to generate the required magnitude of flavor asymmetry of
the sea quark distributions. As a matter of course, a comparison
with the experimental data is premature at this stage of
calculation. First of all, we have not yet paid any
attention to the renormalization scale dependence of the
distribution functions. (We shall later come back to this point.)
Secondly, the distribution functions of Fig.1 obtained within
the valence quark approximation do not saturate the isospin
(or Adler) sum rule. In fact, we obtain
\begin{equation}
   \int_0^1 \,\, {\{ \,[\,u(x) - d(x) \,] \ - \ 
   [\,\bar{u} (x) - \bar{d} (x) \,] \,\}}_{val} \,\,dx 
   \ \ = \ \ 0.775 \,\, ,
\end{equation}
denoting that the remaining $23 \,\%$ of the nucleon isospin
is carried by the Dirac sea quarks.
The isospin sum rule can be made to hold within the valence
quark approximation if the total moment of inertia $I$ in (31) 
is replaced by its valence quark
contribution $I_{val}$, as was done in [18].
However, this would also change all the $O (\Omega^1)$
distribution functions by the factor of $I \,/ \,I_{val} \simeq 1.3$,
which is by no means justifiable.

Anyhow, it is clear that a consistent calculation should also
include the Dirac sea contribution. Since this part is
generally dependent on the selected regularization scheme,
we first investigate in Fig.2 the regularization dependence of
the vacuum polarization contribution to $u(x) - d(x)$ and
$\bar{u} (x) - \bar{d} (x)$. Here, the solid curves are the
results of the Pauli-Villars regularization scheme, while the
dashed and dash-dotted curves are respectively those of the
energy-cutoff scheme with the Gaussian and error-function type
regularization functions.
One sees that the regularization scheme with use of the
energy cutoff has a tendency to suppress distribution functions
at small $x$ as compared with the Pauli-Villars scheme.
However, the differences between the three regularization scheme
are not so large.
One can rather say that the regularization scheme dependence
of the distribution functions is pretty small once the cutoff
parameters are determined so as to reproduce a reasonable value
of the moment of inertia. Note also that
the regularization scheme dependence becomes further
insignificant if we see the sum of the valence and the
vacuum polarization contributions to the structure functions,
since the contribution of the valence level is dominant anyway.
Shown in Fig.3 are the sums of the valence and the vacuum polarization
contributions to $u(x) - d(x)$ and $\bar{u} (x) - \bar{d} (x)$.
Here only the distribution functions obtained with the
Gaussian-type regularization function is shown for the reason
explained above. We have numerically confirmed that the isospin
sum rule (32) is satisfied within the precision of $0.2 \,\%$.
One sees that, as compared with the distribution functions
obtained with the valence quark approximation (Fig.1), a further
enhancement is observed for the excess of the $\bar{d}$ sea over 
the $\bar{u}$ sea inside the proton. 

So far we have postponed the discussion on the energy scale of the
model calculation. Roughly speaking,
the quark distribution functions evaluated here corresponds
to the energy scale of the order of the Pauli-Villars mass
$M_{P.V.} \simeq 600 \,\mbox{MeV}$ or the cutoff energy
$E_{max} \simeq 700 \,\mbox{MeV}$ contained in the Gaussian-form
regularization function.
The $Q^2$-evolution must be taken into account in some way
to be compared with the observed nucleon structure functions
at large $Q^2$.
Fortunately, after elaborate analyses of high $Q^2$ data
by taking account of the perturbative $Q^2$ evolution,
Gl\"{u}ck, Reya and Vogt gave a simple parameterization
of quark distribution functions at a normalization point very
close to the energy scale of the present model [29].
We may therefore carry out a
preliminary comparison of our predictions with their low
scale parameterization of quark distribution functions.
Fig.4 shows this comparison. Here the solid curves stand for the
distribution functions $x \,[ \,u(x) - d(x) \,]$ and
$x \,[\,\bar{u} (x) - \bar{d} (x) \,]$  obtained from the
ones given in Fig.3, whereas the boxes
represent the NLO parameterization of ref.[29].
One can say that the qualitative feature of the NLO parameterization
is nicely reproduced. The main discrepancy between the theory and
the phenomenological fit is that the theoretical distribution
functions have nonzero support beyond $x = 1$.
This unphysical tail of the theoretical distribution functions
comes from an approximate nature of our treatment of the soliton
center-of-mass motion (as well as the collective rotational motion),
which is essentially nonrelativistic.
A simple procedure to remedy this defect was proposed by Jaffe
based on the $1+1$ dimensional bag model [27] and recently
reinvestigated by Gamberg et al. within the context of the NJL
soliton model [28].
(The latter model is essentially equivalent to the CQSM.)
According to the latter authors, the effect of Lorentz
contraction can simply be taken into account by first evaluating
the distribution functions in the soliton rest frame (as we are
doing here) and then by using the simple transformation
\begin{equation}
   f_{IMF} (x) \ \ = \ \ \frac{\Theta (1 - x)}{1 - x} \,\,
   f_{RF} ( \,- \,\ln \,(1 - x) \,) \,\, ,
\end{equation}
as far as the order $\Omega^0$ contributions to the distribution
functions are concerned. We are not sure whether
their proof can be generalized
for the order $\Omega^1$ contributions to distribution functions,
in which we must treat three dimensional collective rotational
motion. Nonetheless, it may be interesting to see the effects of
this transformation, especially on $u(x) - d(x)$.
The distribution functions obtained after this transformation
are shown in Fig.4 by the dashed curves. The crucial influence
of this transformation at large $x$ region is obvious from this
figure. Since (52) is a normalization-preserving transformation,
i.e.
\begin{eqnarray}
   \int_0^\infty \,\,f_{RF} (x) \,\,dx \ \ = \ \ 
   \int_0^1 \,\,f_{IMF} (x) \,\,dx \,\, ,
\end{eqnarray}
it follows that the peaks of the transformed distribution
functions becomes sharper than the original ones.
However, the behavior of the distribution functions at
smaller values of $x$ turns out to be rather insensitive
to this transformation. In this small $x$ region, there
remains a qualitative difference between
the phenomenological distribution functions and the
theoretical ones (obtained after the transformation (52))
especially for the antiquark distribution
$\bar{u} (x) - \bar{d} (x)$.
Note however that the enhanced behavior of the GRV
parameterization at smaller values of $x$
is connected with the Regge-like behavior ($\sim 1 / x^\alpha$
with $0 < \alpha < 1$) of the distribution functions assumed
in their fit. It is clear that such hidden dynamics of the
distribution functions never enters into the model calculations
as carried out here.
Probably, within the model calculation as carried out here,
what is more reliable than the detailed $x$ dependence
would be sum rules obtained after the $x$ integration.
Integrating over $x$, we obtain from Fig.4,
\begin{equation}
   \delta_G \ \ \equiv \ \ \int_0^1 \,\,
   [ \,\bar{u} (x) - \bar{d} (x) \,] \,\,dx \ \ \simeq \ \ 
   - \,0.130 \,\, ,
\end{equation}
or equivalently
\begin{equation}
   S_G \ \ = \ \ \frac{1}{3} \,\,\int_0^1 \,\,
   \{ \,[\,u(x) - d(x) \,] \ + \ 
   [\,\bar{u} (x) - \bar{d} (x) \,] \,\} \,\,dx \ \ \simeq \ \ 
   0.247 \,\, .
\end{equation}
On the other hand, using the NLO parameterization of ref.[29],
we find that
\begin{equation}
   \delta_G \ \ \simeq \ \ - \,0.168 \,\, ,
\end{equation}
or
\begin{equation}
   S_G \ \ \simeq \ \ 0.221 \,\, .
\end{equation}
We can say that the agreement between the theory and the
phenomenological fit is pretty good in view of the extreme
simplicity of the CQSM.
Needless to say, the flavor asymmetric polarization
of the sea quark distributions obtained above would never arise
if there is no flavor asymmetry in the valence quark numbers
of the nucleon.
This then denotes that the NMC observation is nothing
mysterious : it is explained very naturally as a combined
effects of two ingredients, i.e. the apparently existing
flavor asymmetry of the valence quark numbers in the nucleon
and the spontaneous chiral symmetry breaking of the QCD
vacuum [30]. 

Next we turn to the discussion on the flavor-singlet (isoscalar)
polarized distribution function. As has been already pointed out,
the vacuum polarization contribution to this quantity is
convergent without regularization. (This feature comes from
the fact that the flavor-singlet axial-charge is related to the
imaginary part of the Euclidean effective meson
action.) What is more, we find that the Dirac sea contributions
to $\Delta u(x) + \Delta d(x)$ as well as $\Delta \bar{u} (x) + 
\Delta \bar{d} (x)$ are consistent with zero within the numerical
accuracy of the model calculation. This means that the valence
quark approximation is a good approximation for this special
quantity. However, the valence quark approximation here should be
distinguished from the corresponding one in ref.[18] where
the total moment of inertia $I$ had to be replaced by its valence
quark part $I_{val}$ so as to maintain the isospin (or Adler)
sum rule. In making this replacement, one would necessarily
overestimate $O (\Omega^1)$ distribution functions like
the isoscalar polarized one, since they
are inversely proportional to $I$.
In any case, since the vacuum polarization contribution to
$\Delta u(x) + \Delta d(x)$ as well as $\Delta \bar{u} (x) +
\Delta \bar{d} (x)$ are almost negligible, we will
show below only the total contributions.
The solid curves in Fig.5 stand for the isoscalar polarized
distribution functions $\Delta u(x) + \Delta d(x)$ and
$\Delta \bar{u} (x) + \bar{d} (x)$ directly obtained from the
formula (35), whereas the dashed curves here take account of
the effect of Lorentz contraction by making use of the transformation
(52). Again, the sizable effect of Lorentz boost is
self-explanatory especially at larger values of $x$.
Since the low energy scale parameterization
of the polarized distribution functions is not yet available,
we shall postpone the detailed comparison with the observed
structure functions for future studies.
A qualitatively interesting feature observed in
Fig.5 is that the isoscalar combination of $\bar{u}$ and $\bar{d}$
seas is slightly polarized (in most range of $x$) against the
nucleon spin direction. Although this combined polarization of
$\bar{u}$ and $\bar{d}$ seas is rather small in magnitude,
this does not mean that
$\Delta \bar{u} (x)$ and $\Delta \bar{d} (x)$ are both small.
The leading order calculation of Diakonov et al. for the isovector
polarized distribution functions tells us that
$\Delta \bar{u} (x) - \Delta \bar{d} (x)$ is likely to be
positive and fairly large [19,20].
This together with the above result for $\Delta \bar{u} (x) + 
\Delta \bar{d} (x)$ indicates that the seas of
$\bar{u}$ and $\bar{d}$ are sizably
polarized in such a way that their polarizations cancel each other.
Now using the distribution functions given in Fig.5, we can calculate
the first moment of the flavor singlet (longitudinally) polarized
quark distribution functions. We find that
\begin{equation}
   \int_0^1 \,\,[\,\Delta u(x) + \Delta d(x) + 
   \Delta \bar{u} (x) + \Delta \bar{d} (x) \,] \,\,dx \ \ = \ \ 
   0.494 \,\, .
\end{equation}
On the other hand, the direct calculation of the flavor-singlet
axial charge with use of (36) gives
\begin{eqnarray}
   (g_A^{(0)}) \ \ = \ \ 0.492 \,\, .
\end{eqnarray}
The agreement between the above two numbers can be interpreted
as showing the accuracy of our numerical procedure for
evaluating quark distribution functions.
In any case, from the above value of the sum rule,
one reconfirms that the spin fraction of the nucleon carried by quarks
comes out to be quite small (less than half) in the CQSM.
As has been repeatedly
emphasized by one of the authors [16], the
smallness of the quark spin fraction (or the largeness of the
orbital angular momentum) is inseparably connected with the
basic dynamical assumption of this unique model, i.e. the
identification of a rotating hedgehog with the physical nucleon.

\vspace{4mm}
\section{Summary}

\ \ \ In summary, we have evaluated the isovector unpolarized and
isoscalar polarized distribution functions of the nucleon
within the framework of the CQSM with full inclusion of the
Dirac sea quarks . It has been shown that the characteristic features
of the observed distribution functions, i.e. the excess of the
$\bar{d}$ sea over the $\bar{u}$ sea in the proton as well as the
very small quark spin fraction of the nucleon, are reproduced
at least qualitatively within a single theoretical framework
of the CQSM, which is the simplest effective quark model of QCD
taking maximal account of chiral symmetry.
We then interpret this success as revealing the crucial
importance of chiral symmetry in the physics of high energy
deep-inelastic structure functions.

\vspace{5mm}
\section*{Acknowledgement}

 \ \ \ Numerical calculation was performed by using the workstation
``miho'' at the Research Center for Nuclear Physics, Osaka University. 

%
%
\vspace{5mm}
\section*{References}
\newcounter{refnum}
\begin{list}%
{[\arabic{refnum}]}{\usecounter{refnum}}
\item F.E~Close, {\it An Introduction to Quarks and Partons} \ 
(Academic Press, London, 1979).\\
T.~Muta, {\it Foundations of Quantum Chromodynamics} \ 
(World Scientific, Singapore, 1987).
\item EMC Collaboration, J.~Aschman et al., Phys. Lett. 
{\bf B206}, 364 (1988) ; \\
Nucl. Phys. {\bf B328}, 1 (1989).
\item NMC Collaboration, P.~Amaudruz et al.,
Phys. Rev. Lett. {\bf 66}, 2712 (1991).
\item E.M.~Henley and G.A.~Miller, Phys. Lett. {\bf B251}, 453
(1990).
\item S.~Kumano, Phys. Rev. {\bf D43}, 59 (1991) ;\\
S.~Kumano and J.T.~Londergan, Phys. Rev. {\bf D44}, 717 (1991).
\item M.~Wakamatsu, Phys. Rev. {\bf D44}, R2631 (1991) ;
Phys. Lett. {\bf B269}, 394 (1991) ;\\
Phys. Rev. {\bf D46},
3762 (1992).
\item For a recent review, see, e.g. S.~Kumano, hep-ph/9702367.
\item S.J.~Brodsky, J.~Ellis and M.~Karliner, Phys. Lett.
{\bf B206}, 309 (1988).
\item G.~Altarelli and G.G.~Ross, Phys. Lett. {\bf B212},
391 (1988) ;\\
R.~Carlitz, J.C.~Collins and A.~Mueller, Phys. Lett.
{\bf B214}, 229 (1988).
\item S.~Forte and E.V.Syuryak, Nucl. Phys. {\bf B357}, 153
(1991)
\item C.~Best et al., hep-ph/9706502.
\item R.L.~Jaffe, Phys. Rev. {\bf D11}, 1953 (1975).
\item A.I.~Signal and A.W.~Thomas, Phys. Lett. {\bf B211},
481 (1988) ;\\
A.W.~Schreiber, A.I.~Signal and A.W.~Thomas,
Phys. Rev. {\bf D44}, 2653 (1991).
\item V.~Sanjose and V.~Vento, Nucl. Phys. {\bf A501},
672 (1989).
\item D.I.~Diakonov, V.Yu.~Petrov and P.V.~Pobylista, 
Nucl. Phys. {\bf B306}, 809 (1988).
\item M.~Wakamatsu and H.~Yoshiki, Nucl. Phys.
{\bf A524}, 561 (1991).
\item For recent reviews, see, Chr.V.~Christov, A.~Blotz,
H.-C.~Kim, P.~Pobylitsa, T.~Watabe, Th.~Meissner, E.~Ruiz Arriola
and K.~Goeke, Prog. Part. Nucl. Phys. {\bf 37}, 91 (1996) ;\\
R.~Alkofer, H.Reinhardt and H.~Weigel, Phys. Rep. {\bf 265}, 139
(1996).
\item H.~Weigel, L.~Gamberg and H.~Reinhardt, Mod. Phys. Lett.
{\bf A11}, 3021 (1996) ;\\
Phys. Lett. {\bf B399}, 287 (1997).
\item D.I.~Diakonov, V.Yu.~Petrov and P.V.~Pobylista,
M.V.~Polyakov and C.~Weiss,\\
Nucl. Phys. {\bf B480}, 341 (1996). 
\item D.I.~Diakonov, V.Yu.~Petrov and P.V.~Pobylista,
M.V.~Polyakov and C.~Weiss,\\
hep-ph/9703420.
\item K.~Tanikawa and S.~Saito, Nagoya Univ. preprint,
DPNU-96-37 (1996).
\item Chr.V.~Christov, A.~Blotz, K.~Goeke, P.~Pobylitsa, 
V.Yu.~Petrov, M.~Wakamatsu \\
and T.~Watabe, Phys. Lett. {\bf B325}, 467 (1994).
\item M.~Wakamatsu, Prog. Theor. Phys. {\bf 95}, 143 (1996).
\item E.~Braaten, S.-M.~Tse and C.~Willcox, Phys. Rev. {\bf D34},
1482 (1986).
\item S.~Kahana and G.~Ripka, Nucl. Phys. {\bf A429}, 462 (1984).
\item M.~Arneodo et al., Phys. Rev. {\bf D50}, R1 (1994).
\item R.L.~Jaffe, Phys. Lett. {\bf B93}, 313 (1980) ;
Ann. Phys. (NY) {\bf 132}, 32 (1981).
\item L.~Gamberg, H.Reinhardt and H.Weigel, hep-ph/9707352.
\item M.~G\"{u}ck, E.~Reya and A.~Vogt, Z. Phys. {\bf C67},
433 (1995).
\item M.~Wakamatsu, in {\it Weak and Electromagnetic Interactions
in Nucei (WEIN-92)},\\
Proceeding of the International Seminar,
Dubna, Russia, 1992, edited by \\
Ts.D.~Vylov (World Scientific, Singapore, 1993).
\end{list}
\vspace{8mm}
\begin{flushleft}
\large\bf{Figure caption} \\
\end{flushleft}
\ \\
\begin{minipage}{2cm}
Fig. 1.
\end{minipage}
\begin{minipage}[t]{13cm}
The contributions of the valence level to the distribution
functions $u(x) - d(x)$ and $\bar{u} (x) - \bar{d} (x)$.
\end{minipage}
\ \\
\vspace{6mm}
\ \\
\begin{minipage}{2cm}
Fig. 2.
\end{minipage}
\begin{minipage}[t]{13cm}
The contributions of the Dirac continuum to the distribution
functions $u(x) - d(x)$ and $\bar{u} (x) - \bar{d} (x)$.
The solid curves represent the results of the Pauli-Villars
regularization scheme, whereas the dashed and dash-dotted
curves are those of the energy-cutoff scheme with the
Gaussian and error-function type regularization functions.
\end{minipage}
\ \\
\vspace{6mm}
\ \\
\begin{minipage}{2cm}
Fig. 3.
\end{minipage}
\begin{minipage}[t]{13cm}
The sum of the valence and Dirac sea contributions to
$u(x) - d(x)$ and $\bar{u} (x) - \bar{d} (x)$ obtained with the
regularization function of Gaussian form.
\end{minipage}
\ \\
\vspace{6mm}
\ \\
\begin{minipage}{2cm}
Fig. 4.
\end{minipage}
\begin{minipage}[t]{13cm}
The theoretical distribution functions in comparison with the
low energy scale parameterization (NLO fitting) of ref.[29].
The solid curves just correspond to the
distribution functions of Fig.3 multiplied by $x$,
while the dashed curves are obtained
by using the transformation (52).
\end{minipage}
\ \\
\vspace{6mm}
\ \\
\begin{minipage}{2cm}
Fig. 5.
\end{minipage}
\begin{minipage}[t]{13cm}
The isoscalar polarized distribution functions $\Delta u(x) +
\Delta d(x)$ and $\Delta \bar{u} (x) + \Delta \bar{d} (x)$.
The solid curves are directly obtained from (35), while the
dashed curves are obtained with the transformation (52).
\end{minipage}
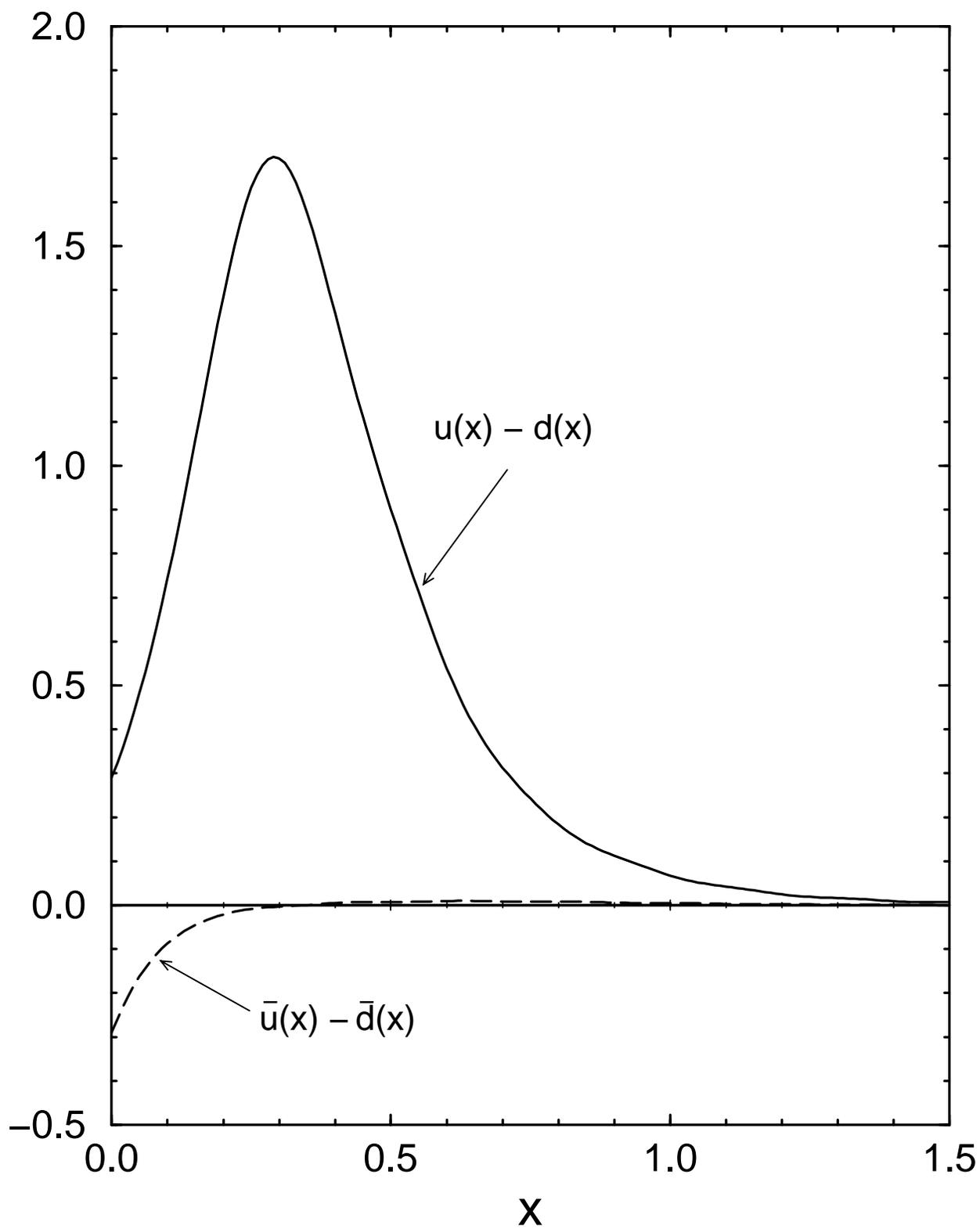
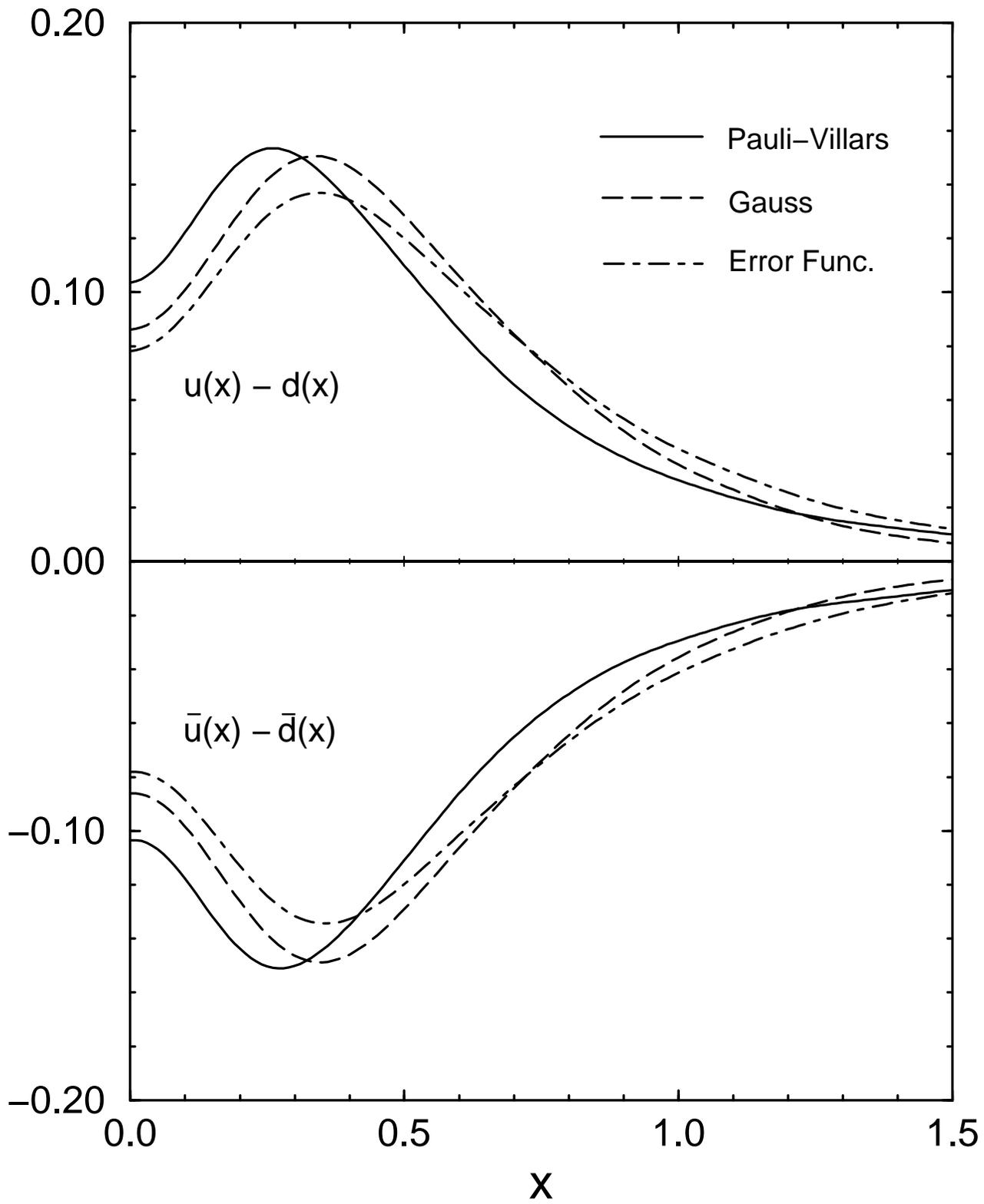
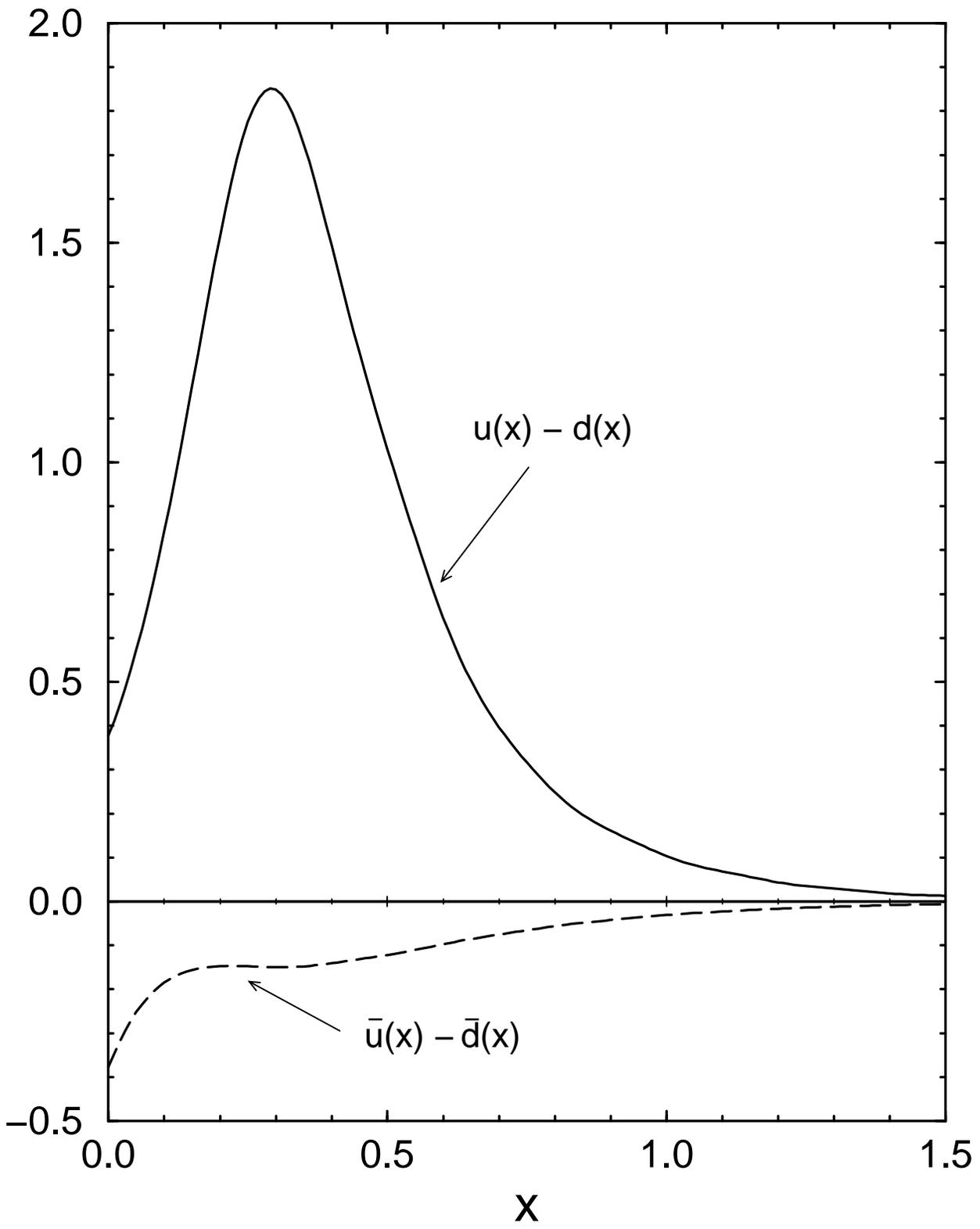
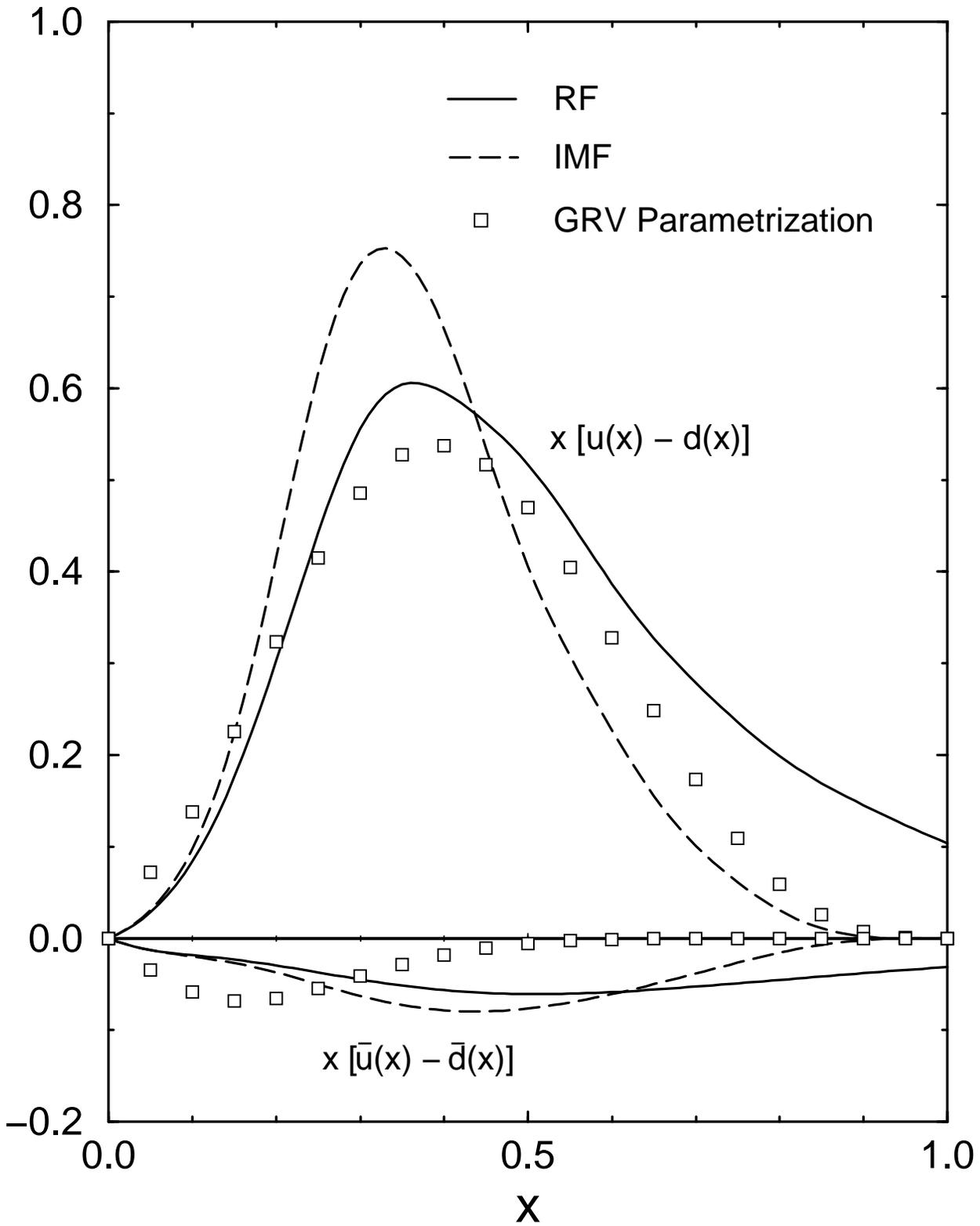
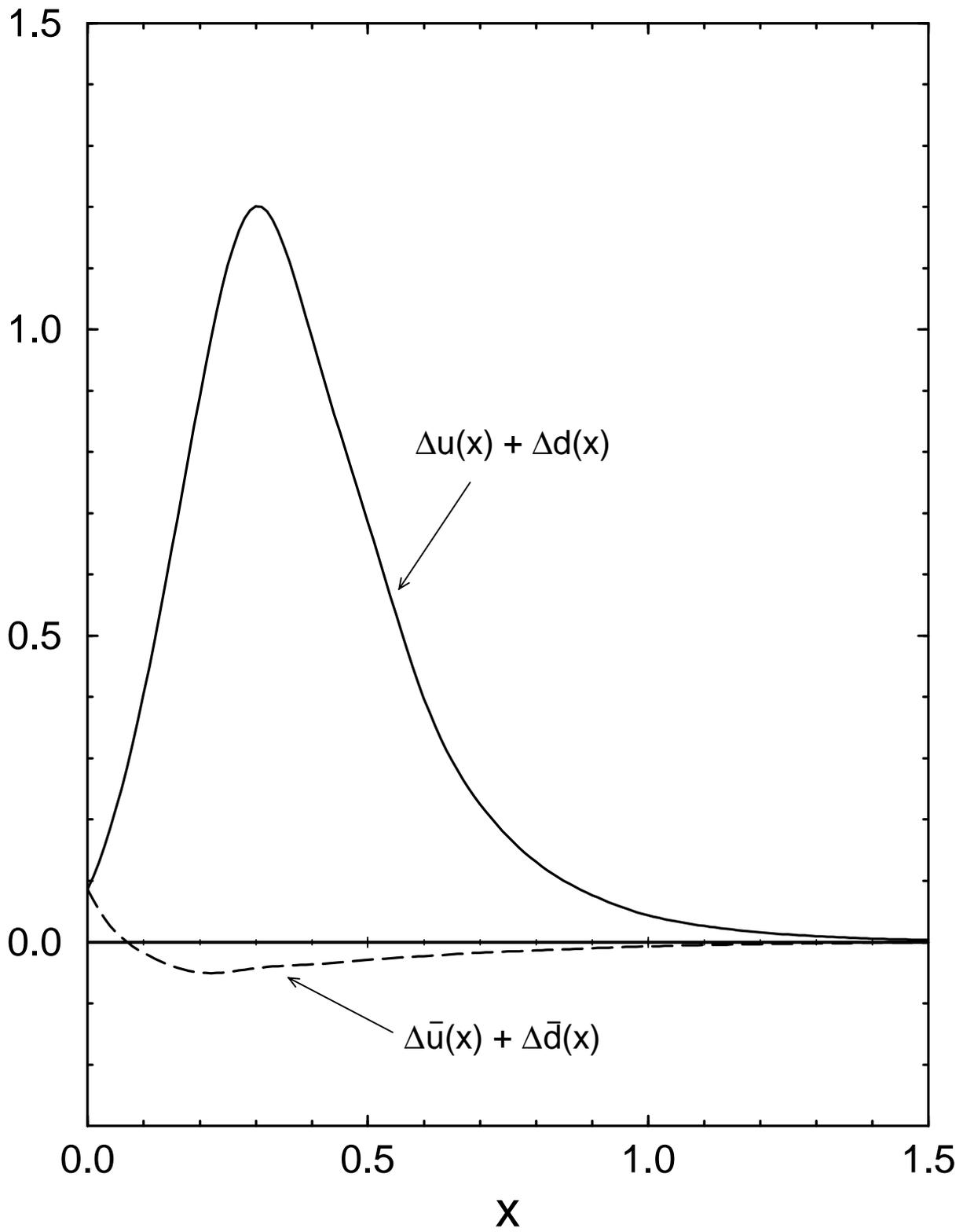
\end{document}